# Performance Comparison of Quasi-Delay-Insensitive Asynchronous Adders

P. Balasubramanian

School of Computer Science and Engineering

Nanyang Technological University

Singapore 639798

## Abstract

In this technical note, we provide a comparison of the design metrics of various quasi-delay-insensitive (QDI) asynchronous adders, where the adders correspond to diverse architectures. QDI adders are robust, and the objective of this technical note is to point to those QDI adders which are suitable for low power/energy and less area. This information could be valuable for a resource-constrained low power VLSI design scenario. Non-QDI adders are excluded from the comparison since they are not robust although they may have optimized design metrics. All the QDI adders were realized using a 32/28nm CMOS process.

## 1. Introduction

The 2017 edition of the International Roadmap for Devices and Systems [1] suggests that asynchronous design could be a potential solution to address the increasing power/energy consumption of a digital circuit or system. Substantiating this, in [2], a 128-point, 16-bit, radix-8 fast Fourier transform (FFT) processor was implemented in the robust QDI asynchronous design style and it was compared with a conventional synchronous FFT processor implementation, and both these were realized using a 65nm CMOS process. It was noted that the QDI FFT processor is 34 times more energy-efficient than its synchronous equivalent. The QDI design style is a promising alternative to the synchronous design style, and different types of QDI implementations exist.

QDI circuits are known to be robust to process, voltage, timing and temperature variations [3,4]. This is important to consider since the issue of variability [5] is commonplace in the nanoelectronics era. Moreover, QDI circuits are less affected by electromagnetic interference compared to synchronous circuits [6]. These properties make QDI circuits preferable for secure applications [7,8]. Further, QDI circuits and systems are modular [9], and hence they are convenient to reuse/replace thus obviating the need for extensive timing re-runs and analysis. Furthermore, QDI circuits are naturally elastic [10] unlike synchronous circuits, and they are suitable for subthreshold operation [11].

A QDI circuit is the practically realizable delay-insensitive circuit which includes the weakest compromise of the isochronic fork [12]. The isochronic fork assumption implies that all the wires branching out from a node/junction would experience concurrent rising or falling signal transitions. Usually, the isochronic fork assumption is confined to a small circuit area and hence their realization is feasible. It has been shown in [13] that QDI circuits are realizable in the nano-electronics regime, which confirms that isochronic forks are physically feasible.

Addition is a fundamental operation in computer arithmetic, which is realized using the adder, and an effective adder design is of interest and importance. *This technical note provides a comparison of the design metrics of several QDI adders, which could be a valuable information for determining which QDI adders are suitable for low power/energy and less area. Note that detailed gate-level circuit schematics for the QDI adders will not be provided in this note since they have been already presented in the literature, and the corresponding literature references will only be cited (in the Tables in Section 4) for the benefit of readers.*

The rest of this technical note is organized as follows. Section 2 gives the nomenclature. Section 3 discusses the design preliminaries of QDI circuits. Section 4 presents the design metrics of various 32-bit QDI adders which correspond to diverse architectures, based on 4-phase return-to-zero (RTZ) and 4-phase return-to-one (RTO) handshaking, and suggests which QDI adders are preferable for low power and less area. Section 5 finally concludes this note.

## 2. Nomenclature

- CLA – Carry Lookahead Adder

- BCLA – Block CLA

- BCLARC – BCLA with Redundant Carry

- BCLG – Block Carry Lookahead Generator

- BCLGRC – BCLG with Redundant Carry

- CCLA – Conventional CLA

- CSLA – Carry Select Adder

- CT – Cycle Time

- PCTP – Power-Cycle Time Product

- RCA – Ripple Carry Adder

- SBFA – Single-Bit Full Adder (which is the conventional full adder)

- DBFA – Dual-Bit Full Adder (i.e., an integration of two SBFAs as one 2-bit adder)

- RCA-SBFA – RCA constructed using SBFAs

- RCA-DBFA – RCA constructed using DBFAs

- Hybrid RCA – RCA constructed using DBFAs and SBFAs

- Hybrid BCLA-RCA – Constructed using a mix of BCLA and RCA-SBFA

- Hybrid BCLARC-RCA – Constructed using a mix of BCLARC and RCA-SBFA

- QDI – Quasi-Delay-Insensitive

- RTZ – Return-To-Zero

- RTO – Return-To-One

# 3. QDI Circuits – Background

The design fundamentals of QDI circuits are discussed here to provide a background.

## 3.1. Data Encoding, Handshaking and Timing Parameters

The general schematic of a QDI circuit stage encompassing delay-insensitive data encoding and a 4-phase handshaking is shown in Figure 1a based on the transmitter-receiver analogy. The corresponding technical schematic is shown in Figure 1b.

In Figure 1b, the current stage and next stage registers are analogous to the transmitter and the receiver shown in Figure 1a, and a QDI circuit is sandwiched between the current stage and the next stage register banks. The register bank comprises a series of registers, with one register allotted for each of the rails of a dual-rail encoded data input. The register refers to a 2-input Muller C-element [14]. The C-element will output 1 or 0 if all its inputs are 1 or 0 respectively. If the inputs to a C-element are not identical then the C-element would retain its existing steady-state. The circles with the marking 'C' represent the C-elements in the figures.

In Figure 1, (X1, X0), (Y1, Y0) and (Z1, Z0) represent the dual-rail encoded primary inputs of the corresponding single-rail inputs X, Y and Z. According to delay-insensitive dual-rail data encoding and the 4-phase RTZ handshaking [9], an input W is encoded as (W1, W0) where W = 1 is represented by W1 = 1 and W0 = 0, and W = 0 is represented by W0 = 1 and W1 = 0. Both these assignments are called data. The assignment W1 = W0 = 0 is called the spacer, and the assignment W1 = W0 = 1 is deemed illegal since the coding scheme should be complete [15] and unordered [16] to maintain the delay-insensitivity.

The application of input data to a QDI circuit which adheres to the 4-phase RTZ handshaking follows the sequence of data-spacer-data-spacer, and so forth. It may be noted that the application of data is followed by the application of the spacer, which implies that there is an interim RTZ phase between the successive applications of input data. The interim RTZ phase ensures a robust data communication (handshaking) between the transmitter and the receiver. The RTZ handshake protocol is specified by the following four steps:

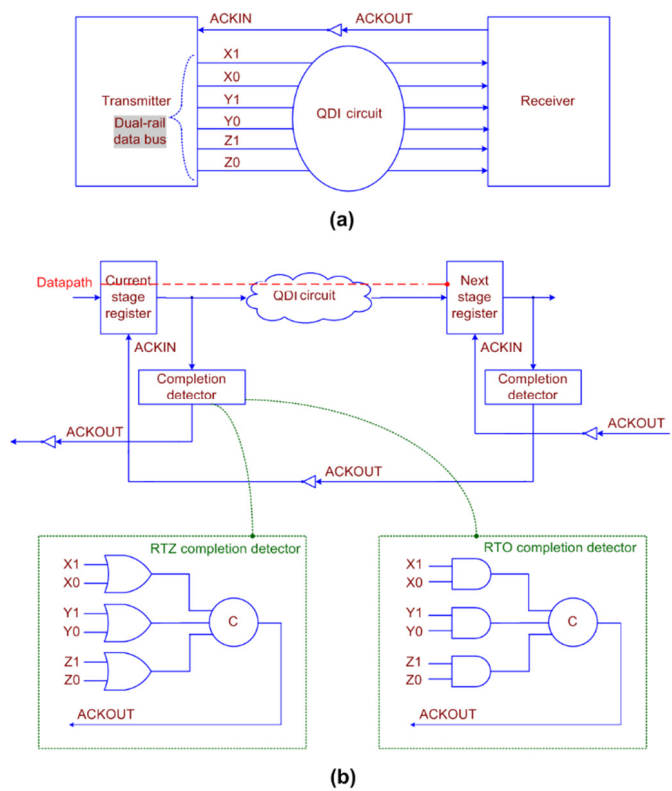

**Figure 1.** (a) Transmitter-Receiver analogy of a QDI circuit stage, and (b) technical schematic portraying the example RTZ and RTO completion detectors for the presumed dual-rail data bus with inputs (X1, X0), (Y1, Y0) and (Z1, Z0). The OR gates and AND gates used in the RTZ and RTO completion detectors are the duals of each other. The datapath is highlighted by the red dashed line in (b).

- First, the dual-rail data bus specified by (X1, X0), (Y1, Y0) and (Z1, Z0) assumes the spacer, and therefore the acknowledgment input (ACKIN) is equal to binary 1. After the transmitter transmits a data, this would cause rising signal transitions i.e., binary 0 to 1 to occur on one of the dual rails of the entire dual-rail data bus

- Second, the receiver would receive the data sent and drive the acknowledgment output (ACKOUT) to 1. ACKIN is the Boolean complement of ACKOUT and vice-versa

- Third, the transmitter waits for ACKIN to become 0 and would subsequently reset the dual-rail data bus, i.e., the dual-rail data bus assumes the spacer again

- Fourth, after an unbounded (but a finite and positive) time duration, the receiver would drive ACKOUT to 0 and then ACKIN would assume 1. With this, a single data transaction is said to be completed and the QDI circuit is permitted to start the next data transaction

According to dual-rail data encoding and the 4-phase RTO handshaking [17], an input V is encoded as (V1, V0) and V = 1 is represented by V1 = 0 and V0 = 1, and V = 0 is represented by V0 = 0 and V1 = 1. Both these assignments are called data. The assignment V1 = V0 = 1 is called the spacer, and the assignment V1 = V0 = 0 is deemed illegal to maintain the delay-insensitivity.

The application of input data to a QDI circuit conforming to the 4-phase RTO handshaking follows the sequence of spacer-data-spacer-data, and so forth. It may be noted that there is an interim RTO phase between the successive applications of input data. The interim RTO phase ensures a proper and robust data communication between the transmitter and the receiver. The RTO handshaking process is specified by the following four steps:

- First, ACKIN is equal to binary 1. After the transmitter transmits the spacer, this would cause rising signal transitions i.e., binary 0 to 1 to occur on all the rails of the entire dual-rail data bus

- Second, the receiver would receive the spacer sent and drive ACKOUT to 1

- Third, the transmitter waits for ACKIN to become 0 and would then transmit the data through the dual-rail data bus

- Fourth, after an unbounded (but a finite and positive) time duration, the receiver would drive ACKOUT to 0 and subsequently ACKIN would assume 1. With this, a single data transaction is said to be completed and the QDI circuit is permitted to start the next data transaction

In a QDI circuit, the time taken to process the data in the datapath, highlighted by the red dashed line in Figure 1b, is called forward latency, and the time taken to process the spacer is called reverse latency. Since there is an intermediate RTZ or RTO phase between the application of two input data sequences, the cycle time (CT) gives the sum of forward and reverse latencies. The CT of a QDI circuit is the equivalent of the clock period of a synchronous circuit. The CT governs the speed at which new data can be input to a QDI circuit.

The gate-level details of example completion detectors corresponding to RTZ and RTO handshaking is shown at the bottom of Figure 1b, within the dotted green boxes. The completion detector indicates i.e., acknowledges the receipt of all the primary inputs given to a QDI circuit stage. In the case of 4-phase RTZ handshaking, ACKOUT is produced using a 2-input OR gate to combine the respective dual rails of each encoded primary input and then synchronizing the outputs of all the 2-input OR gates using a C-element or a tree of C-elements. In the case of 4-phase RTO handshaking, ACKOUT is produced using a 2-input AND gate to combine the respective dual rails of each encoded primary input and subsequently synchronizing the outputs of all the 2-input AND gates using a C-element or a tree of C-elements.

## 3.2. QDI Circuits

QDI circuits are robust and are classified into three types as strong-indication [18,19], weak-indication [18,20], and early output [21] circuits. The input-output timing relations of QDI circuits are illustrated by the representative timing diagrams in Figures 2a and 2b with respect to RTZ and RTO handshaking.

Strong-indication circuits would wait to receive all the primary inputs (data and spacer), and after receiving them would process them to produce the required primary outputs (data and spacer respectively). On the other hand, weak-indication circuits can produce all but one of the primary outputs after receiving a subset of the primary inputs. Nevertheless, only after receiving the last primary input, they would produce the last primary output. Weak-indication may be enabled locally or globally, and it has been shown in [22,23] that local weak-indication is preferable compared to global weak-indication for QDI function blocks. The weak-indication QDI adders considered in this work adopt local weak-indication.

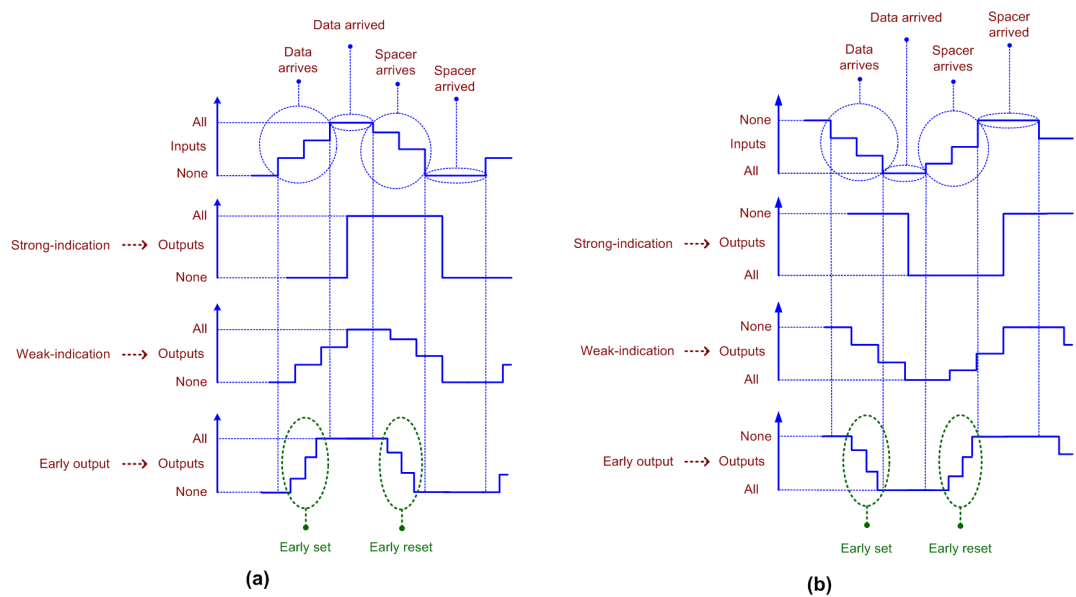

**Figure 2.** Input-output timing relation of different types of QDI circuits corresponding to (a) RTZ handshaking, and (b) RTO handshaking. The early set and reset behaviours of early output circuits are highlighted by the dotted green ovals in (a) and (b).

A connection of strong-indication sub-circuits may not result in a strong-indication circuit; rather, a weak-indication circuit may result. For example, if two strong-indication full adders are connected, it could result in a weak-indication 2-bit RCA. This is because if all the inputs to one of the full adders are provided, the corresponding sum and carry output bits of that full adder could be produced regardless of the non-arrival of inputs to the other full adder in the RCA. However, only after all the inputs to the other full adder are provided, its corresponding sum and carry output bits would be produced. This scenario is characteristic of weak-indication.

While comparing strong- and weak-indication circuit types, the latter are preferable [24,25], and this is because of the strict timing restrictions inherent in the former. Especially, for implementing arithmetic functions, the weak-indication type is preferable to the strong-indication type and this is due to the following reasons: i) strong-indication arithmetic circuits tend to encounter worst-case forward and reverse latencies for the application of data and spacer, and therefore the CT of strong-indication arithmetic circuits is always the maximum (i.e., worst-case timing), ii) weak-indication arithmetic circuits may encounter data-dependent forward and reverse latencies or a data-dependent forward latency and a constant reverse latency, and so the CTs of weak-indication arithmetic circuits are usually less compared to those of strong-indication arithmetic circuits.

An early output circuit is however more relaxed compared to the strong- and weak-indication counterparts. After receiving a subset of the primary inputs (data or spacer), an early output circuit can produce all the primary outputs (data or spacer respectively). This implies the late arriving primary inputs may not be acknowledged by the circuit. However, this does not cause any concern because isochronic fork assumptions are imposed on all the primary inputs, and all the primary inputs are provided to the completion detector that precedes the early output circuit, as seen in Figure 1b. Hence, the acknowledgment of the late arriving primary inputs by the completion detector also implies the receipt of those primary inputs by the QDI circuit. Thus, the problem of wire orphan(s) i.e., unacknowledged signal transitions on the wire(s) due to the late arriving input(s) is overcome through the assumption of isochronic forks which is imposed on all the primary inputs.

Either the data may be produced early, or the spacer may be produced early in an early output circuit. Accordingly, an early output circuit is categorized as early set or early reset kind. The early set and early reset behaviours of early output circuits are highlighted by the dotted green ovals in Figures 2a and 2b. An early output RCA is preferable to a strong-indication and a weak-indication RCA for achieving improved optimizations in speed and power/energy. In general, an early output circuit can achieve enhanced optimizations in the design metrics compared to the strong- and weak-indication counterparts.

In a QDI circuit, the logic decomposition should be performed safely [26,27]. Although many logic decomposition (factorization) techniques exist [28], safe QDI logic decomposition is essential to avoid the problem of gate orphans, which are unacknowledged signal transitions occurring on the intermediate gate output(s). For an illustration of gate and wire orphans, the interested reader is referred to [29-31].

The signal transitions will have to occur monotonically throughout an entire QDI circuit from the first logic level, which receives the primary inputs, up to the last logic level, which produces the primary outputs [32]. The signal transitions should either be seen as rising or falling throughout an entire QDI circuit. In general, the signal transitions will be rising (i.e., binary 0 to 1) for the application of data and falling (i.e., binary 1 to 0) for the application of spacer in a QDI circuit that corresponds to RTZ handshaking. On the other hand, the signal transitions will be rising for the application of spacer and falling for the application of data in a QDI circuit that corresponds to RTO handshaking.

For monotonicity of signal transitions, the monotonic cover constraint [9] should be incorporated into a QDI logic description. For example, this implies if a QDI logic function is expressed in the sum-of-products form, only one product term should be activated for the application of an input data, i.e., the product terms comprising the sum-of-products expression of a QDI logic function should be mutually orthogonal (also called disjoint), and the logical conjunction of any two product terms in a QDI logic function should yield zero. Thus, a QDI logic function is ideally expressed in the disjoint sum-of-products form [33-35], which would consist of mutually disjoint product terms to satisfy the monotonic cover constraint. An example illustration of the monotonic cover constraint is given in Section 2.2 of [36], and an interested reader may refer to the same. Embedding the monotonic cover constraint and performing safe QDI logic decomposition are vital to the correct implementation of a QDI circuit.

Incorporating the monotonic cover constraint in a QDI logic function would cause the activation of just one signal propagation path from a primary input to a primary output for the application of input data. This is useful to facilitate the proper acknowledgment of signal transitions throughout an entire QDI circuit, thus avoiding the likelihood of any gate orphan occurrence(s). Gate orphans are troublesome unlike wire orphans as they may affect the robustness of a QDI circuit and if they are imminent, restricting them from affecting the circuit robustness may require incorporating additional timing assumptions which are likely to be sophisticated and may also be practically difficult to realize [37].

## 4. Design Metrics of QDI Adders

Several 32-bit QDI adders, which correspond to the generic architectures such as RCA, CSLA, CCLA and BCLA, were physically realized using a 32/28nm CMOS technology [38], corresponding to both RTZ and RTO handshaking. To transform a QDI circuit corresponding to RTZ handshaking into one that corresponds to RTO handshaking and vice-versa, some rules have been defined in [58], and the proofs for these are given in [59]. The 2-input C-element was alone custom-realized by modifying the AO222 gate to implement the QDI adders. A typical-case PVT specification of the high $V_t$ standard digital cell library with a supply voltage of 1.05V and an operating junction temperature of 25°C was considered for the implementations and simulations. The registers and completion detectors associated with the QDI adders are maintained the same with respect to RTZ and RTO handshaking, separately. This implies that the differences between the simulation results of the QDI adders are attributable to the differences between their logic compositions.

About 2000 (random) input vectors encompassing data and spacer, which separately correspond to RTZ and RTO handshaking were used to verify the functionalities of the adders. The input vectors corresponding to RTZ and RTO handshaking bear a logical equivalence. The functional simulations of the QDI adders were successfully performed and their respective switching activities were captured, which were subsequently used to estimate the average power dissipation. Synopsys EDA tools were used to estimate the design metrics of the adders. Default wire loads were automatically included while performing the simulations. A virtual clock was used to constrain the input and output ports of the QDI adders, and it did not consume any power.

The design metrics estimated include forward and reverse latencies, CT, area, and average power dissipation. The forward latency of a QDI circuit is similar to the critical path delay of a synchronous circuit and it is directly estimated. The reverse latencies of some QDI adders may differ from their forward latencies. This may be evident from Figure 3. The reverse latencies of QDI adders were estimated from the gate-level simulation timing data, and this method was followed for RTZ and RTO handshaking.

The estimated design metrics of various QDI adders corresponding to RTZ handshaking are given in Table 1, and the design metrics corresponding to RTO handshaking are given in Table 2. Adder legends are provided in the second column of Tables 1 and 2 to help with the discussion. The related literature references pertaining to the QDI adders are given in Tables 1 and 2. RCAs (i.e. RCA-SBFAs) utilising the early output full adders of [40-42] are excluded from the comparison since these RCAs are relative-timed [43]. Relative-timed asynchronous circuits are not QDI and they are non-robust since they usually incorporate additional timing assumptions with respect to sequencing the arrival of internal signals besides the assumption of isochronic forks, and such additional timing assumptions may not be easy to realize. Also, the dual-bit full adders (DBFAs) proposed in our earlier works [56] and [57], which incorporate dual-rail and dual-rail-cum-1-of-4 encodings, are not considered for comparison here since the DBFAs proposed in [49-51] have improved design metrics compared to those.

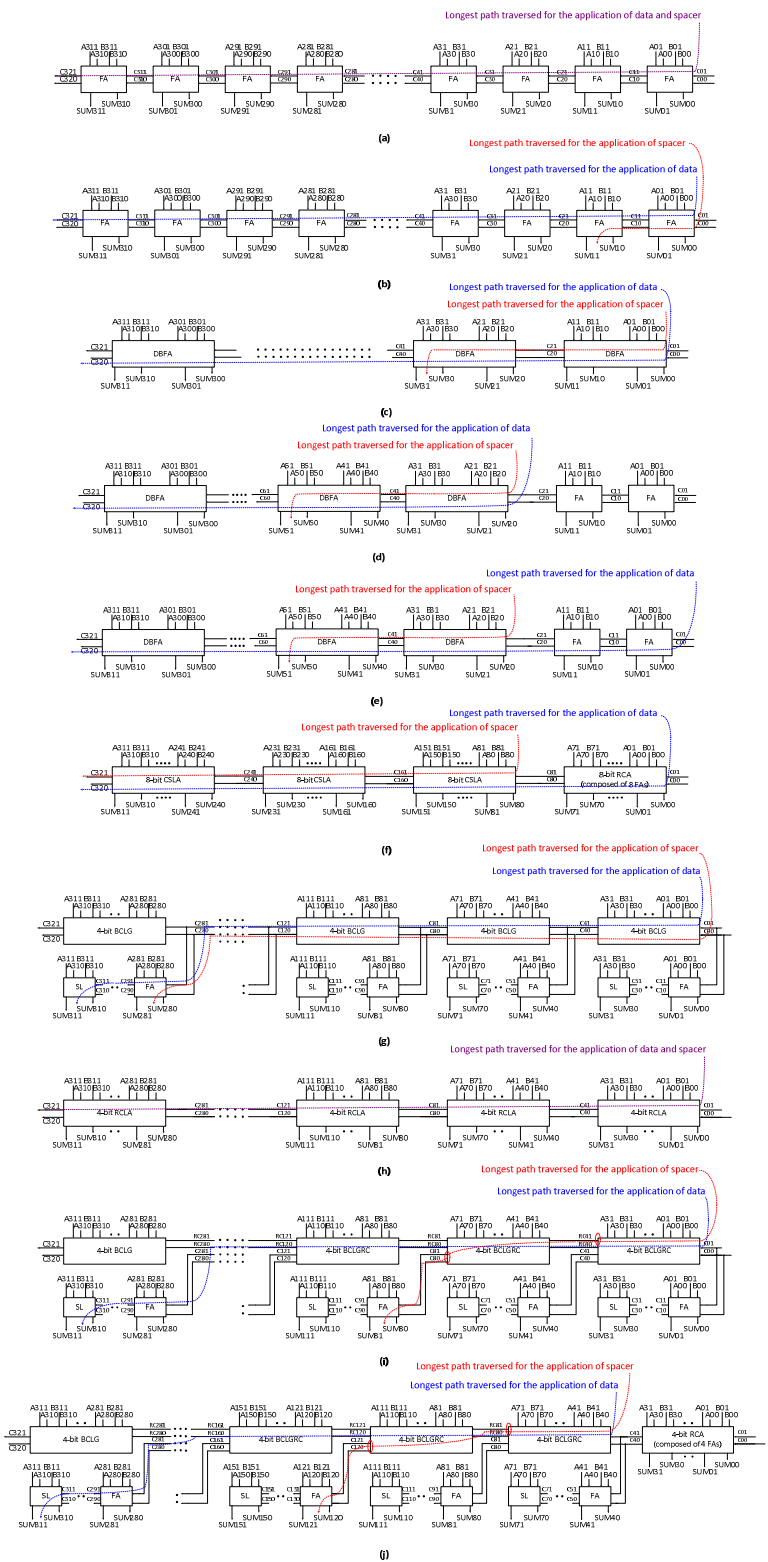

**Figure 3.** Longest signal propagation paths of QDI adders for application of data and spacer. FA refers to the Full Adder and SL refers to the Sum Logic (i.e., FA without the carry output).

**Table 1.** Design metrics of various 32-bit QDI adders based on RTZ handshaking.

| Adder Architecture | Adder Legend | Literature Reference | FL[1] (ns) | RL[2] (ns) | CT[3] (ns) | Area ($\mu m^2$) | Power ($\mu W$) |
|---|---|---|---|---|---|---|---|
| RCA-SBFA | Z1 | [44] | 14.61 | 14.61 | 29.22 | 2529.00 | 2190 |
| RCA-SBFA | Z2 | [45][4] | 9.26 | 9.26 | 18.52 | 2504.60 | 2181 |
| RCA-SBFA | Z3 | [26] | 9.04 | 9.04 | 18.08 | 2293.14 | 2172 |
| RCA-SBFA | Z4 | [45][5] | 8.24 | 8.24 | 16.48 | 2423.27 | 2177 |
| RCA-SBFA | Z5 | [46] | 7.00 | 7.00 | 14.00 | 2016.63 | 2171 |
| RCA-SBFA | Z6 | [47] | 4.43 | 0.58 | 5.01 | 2097.96 | 2174 |
| RCA-SBFA | Z7 | [48] | 3.32 | 0.73 | 4.05 | 2049.16 | 2171 |
| RCA-SBFA | Z8 | [21] | 3.10 | 0.61 | 3.71 | 1658.80 | 2161 |
| RCA-DBFA | Z9 | [49] | 2.23 | 0.78 | 3.01 | 2488.33 | 2173 |
| Hybrid RCA | Z10 | [50] | 2.16 | 0.78 | 2.94 | 2436.48 | 2173 |
| RCA-DBFA | Z11 | [51] | 2.19 | 0.93 | 3.12 | 2000.36 | 2183 |
| Hybrid RCA | Z12 | [51] | 2.19 | 0.93 | 3.12 | 1979.01 | 2182 |
| Uniform CSLA | Z13 | | 2.46 | 1.89 | 4.35 | 3000.17 | 2293 |
| Non-uniform CSLA | Z14 | [52] | 3.23 | 3.23 | 6.46 | 3384.44 | 2312 |
| BCLA | Z15 | | 3.31 | 2.93 | 6.24 | 2951.88 | 2191 |
| BCLARC | Z16 | [53][6] | 2.46 | 1.69 | 4.15 | 2987.46 | 2192 |
| BCLA | Z17 | | 3.14 | 2.88 | 6.02 | 2915.29 | 2188 |
| BCLARC | Z18 | [53][7] | 2.32 | 1.68 | 4.00 | 2950.87 | 2189 |
| CCLA | Z19 | [54] | 2.75 | 2.75 | 5.50 | 2569.65 | 2177 |
| BCLA | Z20 | | 3.13 | 2.88 | 6.01 | 2524.92 | 2178 |
| BCLARC | Z21 | [55] | 2.31 | 1.67 | 3.98 | 2560.50 | 2179 |
| BCLA | Z22 | | 2.76 | 2.50 | 5.26 | 2209.78 | 2174 |
| BCLARC | Z23 | | 2.01 | 1.38 | 3.39 | 2245.36 | 2176 |
| Hybrid BCLARC-RCA1 | Z24 | [36] | 1.93 | 1.38 | 3.31 | 2171.41 | 2174 |
| Hybrid BCLARC-RCA1 | Z25 | | 1.97 | 1.38 | 3.35 | 2097.45 | 2172 |
| Hybrid BCLARC-RCA1 | Z26 | | 2.23 | 1.38 | 3.61 | 2023.49 | 2170 |
| BCLA | Z27 | | 3.46 | 3.20 | 6.66 | 2307.37 | 2187 |
| BCLARC | Z28 | | 1.76 | 1.11 | 2.87 | 2342.95 | 2188 |
| Hybrid BCLARC-RCA1 | Z29 | [39] | 1.86 | 1.11 | 2.97 | 2256.80 | 2184 |
| Hybrid BCLARC-RCA2 | Z30 | | 2.11 | 1.11 | 3.22 | 2170.64 | 2181 |
| Hybrid BCLARC-RCA3 | Z31 | | 2.36 | 1.11 | 3.47 | 2084.49 | 2178 |

[1]Forward Latency; [2]Reverse Latency; [3]Cycle Time; [4]Uses the strong-indication full adder; [5]Uses the weak-indication full adder; [6]Uses the full adder of [47]; [7]Uses the full adder of [48].

**Table 2.** Design metrics of various 32-bit QDI adders based on RTO handshaking.

| Adder Architecture | Adder Legend | Literature Reference | FL[1] (ns) | RL[2] (ns) | CT[3] (ns) | Area ($\mu$m²) | Power ($\mu$W) |
|---|---|---|---|---|---|---|---|
| RCA-SBFA | O1 | [44] | 14.15 | 14.15 | 28.30 | 2529.00 | 2185 |
| RCA-SBFA | O2 | [45][4] | 8.74 | 8.74 | 17.48 | 2374.48 | 2167 |
| RCA-SBFA | O3 | [26] | 8.88 | 8.88 | 17.76 | 2293.15 | 2168 |
| RCA-SBFA | O4 | [45][5] | 8.03 | 8.03 | 16.06 | 2358.21 | 2167 |
| RCA-SBFA | O5 | [46] | 6.95 | 6.95 | 13.90 | 2016.63 | 2167 |
| RCA-SBFA | O6 | [47] | 3.79 | 0.56 | 4.35 | 2097.96 | 2170 |
| RCA-SBFA | O7 | [48] | 3.31 | 0.72 | 4.03 | 2049.16 | 2167 |
| RCA-SBFA | O8 | [21] | 2.93 | 0.61 | 3.54 | 1658.80 | 2157 |
| RCA-DBFA | O9 | [49] | 2.23 | 0.79 | 3.02 | 2716.07 | 2177 |
| Hybrid RCA | O10 | [50] | 2.16 | 0.79 | 2.95 | 2649.97 | 2176 |
| RCA-DBFA | O11 | [51] | 2.17 | 0.91 | 3.08 | 2000.36 | 2179 |
| Hybrid RCA | O12 | [51] | 2.19 | 0.91 | 3.10 | 1979.01 | 2177 |
| Uniform CSLA | O13 | [52] | 2.38 | 1.85 | 4.23 | 3000.17 | 2285 |
| Non-uniform CSLA | O14 |  | 3.15 | 3.08 | 6.23 | 3384.44 | 2303 |
| BCLA | O15 | [53][6] | 3.19 | 2.86 | 6.05 | 2984.41 | 2184 |
| BCLARC | O16 |  | 2.36 | 1.69 | 4.05 | 3019.99 | 2185 |
| BCLA | O17 | [53][7] | 3.10 | 2.84 | 5.94 | 2947.82 | 2182 |
| BCLARC | O18 |  | 2.30 | 1.67 | 3.97 | 2983.40 | 2183 |
| CCLA | O19 | [54] | 2.73 | 2.73 | 5.46 | 2553.39 | 2169 |
| BCLA | O20 | [55] | 3.06 | 2.76 | 5.82 | 2557.45 | 2171 |
| BCLARC | O21 |  | 2.26 | 1.66 | 3.92 | 2593.03 | 2172 |
| BCLA | O22 | [36] | 2.73 | 2.50 | 5.23 | 2193.52 | 2167 |
| BCLARC | O23 |  | 1.95 | 1.37 | 3.32 | 2229.10 | 2168 |
| Hybrid BCLARC-RCA1 | O24 |  | 1.88 | 1.37 | 3.25 | 2157.17 | 2167 |
| Hybrid BCLARC-RCA1 | O25 |  | 1.89 | 1.37 | 3.26 | 2085.25 | 2165 |
| Hybrid BCLARC-RCA1 | O26 |  | 2.13 | 1.37 | 3.50 | 2013.33 | 2164 |
| BCLA | O27 | [39] | 3.38 | 3.14 | 6.52 | 2315.51 | 2180 |
| BCLARC | O28 |  | 1.74 | 1.15 | 2.89 | 2351.09 | 2181 |
| Hybrid BCLARC-RCA1 | O29 |  | 1.78 | 1.15 | 2.93 | 2263.92 | 2178 |
| Hybrid BCLARC-RCA2 | O30 |  | 2.02 | 1.15 | 3.17 | 2176.74 | 2175 |
| Hybrid BCLARC-RCA3 | O31 |  | 2.26 | 1.15 | 3.41 | 2089.57 | 2172 |

[1]Forward Latency; [2]Reverse Latency; [3]Cycle Time; [4]Uses the strong-indication full adder; [5]Uses the weak-indication full adder; [6]Uses the full adder of [47]; [7]Uses the full adder of [48]

Referring to Tables 1 and 2, Z1 (O1) is an RCA constructed using the strong-indication full adder of [44], Z2 (O2) is an RCA constructed using the strong-indication full adder of [45] and Z3 (O3) is an RCA constructed using the strong-indication full adder of [26]. Z4 (O4) is an RCA constructed using the weak-indication full adder of [45], and Z5 (O5) is an RCA constructed using the weak-indication full adder of [46]. The (worst-case) forward and reverse latencies of Z1(O1), Z2 (O2), Z3(O3), Z4 (O4) and Z5 (O5) are governed by the longest signal propagation path shown in violet in Figure 3a. Z6 (O6), Z7 (O7), Z8 (O8) are RCAs which are constructed using the weak-indication full adders of [47] and [48] and the early output full adder of [21] respectively. The forward and reverse latencies of Z6 (O6), Z7 (O7) and Z8 (O8) are governed by the signal propagation paths highlighted in blue and red respectively in Figure 3b. Note that in the case of Z6 (O6), Z7 (O7) and Z8 (O8), their reverse latency is a constant, which is typically governed by two full adder stages, while their forward latency is input data-dependent. Z9 (O9) and Z11 (O11) are 32-bit RCAs, constructed using 16 early output dual-bit full adders (DBFAs) of [49] and [51] respectively, and their forward and reverse latencies are governed by the signal propagation paths shown in blue and red in Figure 3c. Z10 (O10) of [50] is an improved, i.e., a hybrid RCA version of Z9 (O9) in that a 2-bit least significant RCA comprising two single-bit full adders (SBFAs) of [21] are used to replace a least significant DBFA of [49]. The resultant is a reduction in the forward latency, as depicted by the datapath shown in blue in Figure 3d, while the reverse latency is the same as Figure 3c. Z12 (O12) is similarly an improved, i.e., a hybrid RCA version of Z11 (O11) in that a 2-bit least significant RCA comprising two single-bit full adders (SBFAs) of [21] are used to replace a least significant DBFA of [51]. However, this does not result in a reduction of the forward latency, and in fact the forward latency of Z12 (O12) is slightly greater than the forward latency of Z11 (O11), while their reverse latencies are the same. This implies that a hybrid RCA may not always lead to a reduction in the CT compared to a regular RCA. This was also noticed when comparing the hybrid BCLARC-RCAs with the BCLARCs of [39], where the latter is beneficial than the former in terms of the CT.

Z13 (O13) and Z14 (O14) are non-uniform input-partitioned and uniform input-partitioned CSLAs presented in [52]. These CSLAs were constructed using the early output full adder of [21] and a strong-indication 2:1 multiplexer of [60]. Of the two CSLAs, the 32-bit CSLA based on an 8-8-8-8 uniform input-partition was found to have better design metrics compared to the non-uniform input-partitioned CSLA. The longest signal propagation paths corresponding to the forward and reverse latencies are shown in blue and red in Figure 3f. The internal details of the 8-bit sub-CSLA and the 8-bit RCA are given in [52].

Z15 (O15) and Z17 (O17) are BCLAs, which utilize the BCLG proposed in [53]. While Z15 (O15) utilizes the weak-indication full adder of [47], Z17 (O17) utilizes the latency optimized weak-indication full adder of [48]. Z16 (O16) and Z18 (O18) are derived from Z15 (O15) and Z17 (O17) respectively, which are BCLARCs. It has been observed in [61] that incorporating redundant logic especially for the carry propagation logic within a QDI adder could help to reduce the latencies and also the CT. As a result, the BCLARCs feature a significant reduction in the latencies and CT compared to the BCLAs. This observation holds good for a comparison between Z20 (O20) and Z21 (O21) of [55], Z22 (O22) and Z23 (O23) of [36], and Z27 (O27) and Z28 (O28) of [39]. The longest signal propagation paths corresponding to forward and reverse latencies in the case of a BCLA are represented by the dotted blue and red lines in Figure 3g, while the longest signal propagation paths corresponding to forward and reverse latencies in the case of a BCLARC are represented by the dotted blue and red lines in Figure 3i. By comparing Figure 3g and 3i, it is clear that the reverse latency of a BCLARC is much reduced than the reverse latency of a BCLA and this is because of the introduction of the redundant carry output logic.

A conventional QDI CCLA was presented in [54], which is represented as Z19 and O19 in Tables 1 and 2. While the forward latency of the CCLA is less compared to the forward latencies of BCLAs, unfortunately, the reverse latency of the CCLA is the same as the forward latency, as seen from Figure 3h. As a result, the CT of the CCLA is greater than the CTs of BCLAs, BCLARCs and hybrid BCLARC-RCAs.

To further reduce the forward latency of BCLARCs, a small-size RCA may be of use in the least significant adder bit positions. This leads to a hybrid BCLARC-RCA architecture, which is represented by Z24 (O24) to Z26 (O26) and Z29 (O29) to Z31 (O31) in Tables 1 and 2. Z24 (O24), Z25 (O25) and Z26 (O26) embed a least significant 4-bit, 8-bit and 12-bit RCA. Likewise, Z29 (O29), Z30 (O30) and Z31 (O31) embed a least significant 4-bit, 8-bit and 12-bit RCA. However, the usefulness or no use of the hybrid BCLARC-RCA architecture has to be verified via timing analysis. In the case of the BCLARC of [36], the replacement of a least significant 4-bit BCLARC by a 4-bit RCA enables a small reduction in the forward latency and also the CT (the characteristic of which is portrayed by Figure 3j), while the use of a higher order RCA is found to degrade the forward latency. However, in the case of the BCLARC proposed in [39], a hybrid BCLARC-RCA configuration does not lead to any improvement in the CT.

Overall, when considering all the QDI adders given in Tables 1 and 2, it becomes clear that in terms of the CT, the BCLARC proposed in our latest work [39] (represented by Z28 and O28) is better optimized compared to the rest.

The CT governs the speed of a QDI circuit that utilizes delay-insensitive data encoding and a 4-phase handshaking, and the power-cycle time product (PCTP) governs the low power/low energy aspect. Hence, the PCTPs of the QDI adders were calculated and then normalized. The normalization was performed such that the highest PCTP among the set of QDI adders corresponding to a handshake protocol was normalized to 1, and the actual PCTPs of the remaining adders were divided by the highest PCTP. Thus, after normalization, the least value of the PCTP reflects the optimum low power/energy design. The plots of normalized CT and PCTP values corresponding to RTZ handshaking are shown side-by-side in Figures 4a and 4b, and the similar plots for RTO handshaking are shown in Figures 5a and 5b.

CT predominantly influences the PCTP of QDI adders. This is because the average power dissipations of QDI adders are quite the same and this is because all the QDI adders satisfy the monotonic cover constraint [9]. The average power of the QDI adders are confined to small ranges of 151μW (i.e., 2161μW to 2312μW) in the case of RTZ handshaking and 146μW (i.e., 2157μW to 2303μW) in the case of RTO handshaking. Hence, the PCTP is quite a reflection of CT, as evident from the curves in Figures 4a and 4b, and Figures 5a and 5b. The normalized PCTP plots reveal that Z28 (O28) of [39] is energy-efficient than the rest, as was found to be the case with CT from Tables 1 and 2.

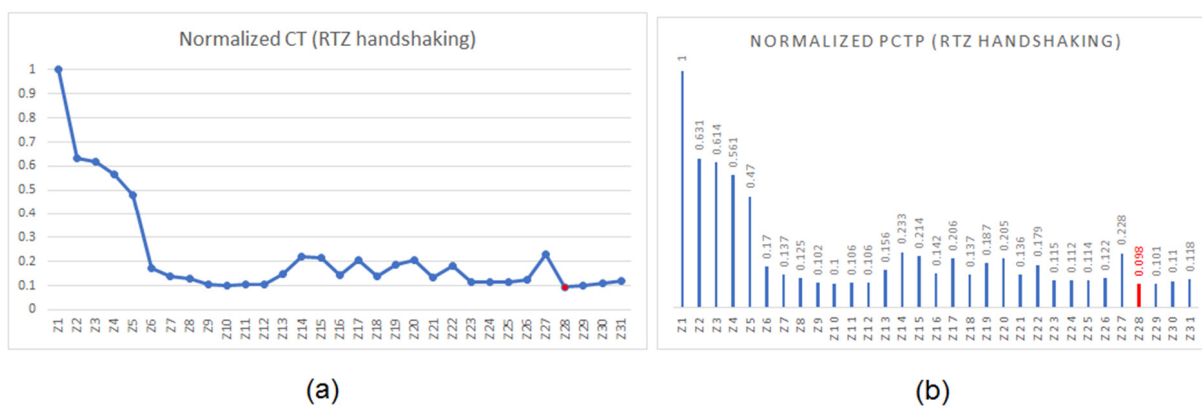

(a)                                        (b)

**Figure 4.** Plots of normalized values of (a) CT and (b) PCTP of 32-bit QDI adders corresponding to RTZ handshaking. The adder legends are referenced from Table 1. The red bar in (b) corresponds to the proposed BCLARC of [39] which is energy-efficient than the rest.

Lastly, in terms of area, Z8 (O8), which is based on the early output full adder of [21], occupies relatively less silicon and dissipates less average power compared to the rest. The full adder of [21] requires less area compared to the full adders of [26,44,45,46,47,48]. Even with respect to a synchronous design the RCA architecture occupies less area and dissipates less power than the other adder architectures [62,63], and this is found to hold good for a QDI design. However, in terms of the CT, Z8 (O8) is 29.3% (22.5%) more expensive than Z28 (O28). As a result, in terms of the energy (PCTP), Z8 (O8) is 27.7% (21.1%) more expensive compared to Z28 (O28).

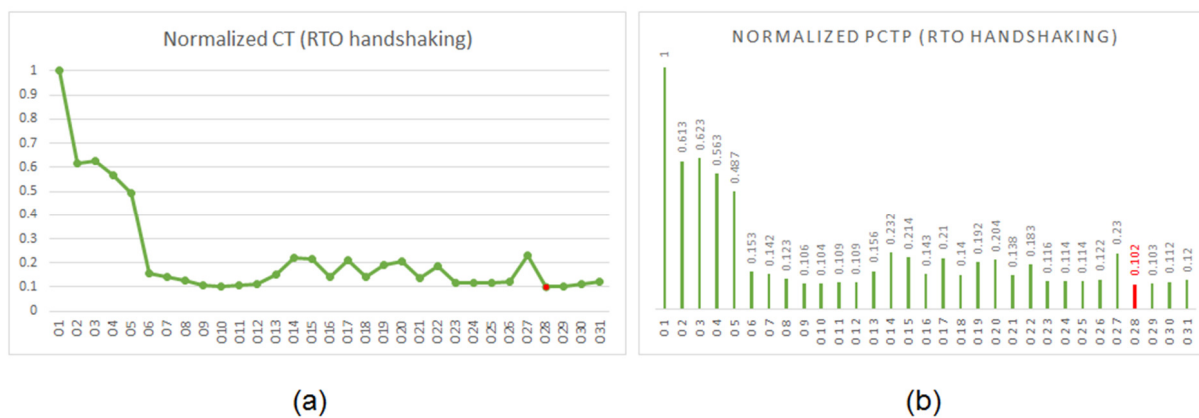

**Figure 5.** Plots of normalized values of (a) CT and (b) PCTP of 32-bit QDI adders corresponding to RTO handshaking. The adder legends are referenced from Table 2. The red bar in (b) corresponds to the proposed BCLARC of [39] which is energy-efficient than the rest.

## 5. Conclusions

This technical note has summarized the design metrics of various QDI adders presented in the literature by considering an example 32-bit addition. Area is not that much of a concern as speed and energy. However, if area becomes an overarching concern, then Z8 (O8) of [21] is preferable. Nevertheless, it was observed that Z28 (O28) of [39] is preferable for implementing high-speed and energy-efficient QDI asynchronous addition based on RTZ (RTO) handshaking.

## References

1. Available: https://irds.ieee.org/roadmap-2017 (last accessed on 19 November 2018).

2. B.Z. Tang and F. Lane, "Low power QDI asynchronous FFT," *Proceedings of 22nd IEEE International Symposium on Asynchronous Circuits and Systems*, pp. 87-88, 2016.

3. C.H. van Berkel, M.B. Josephs and S.M. Nowick, "Applications of asynchronous circuits," *Proceedings of the IEEE*, vol. 87, no. 2, pp. 223-233, 1999.

4. A.J. Martin and M. Nystrom, "Asynchronous techniques for system-on-chip design," *Proceedings of the IEEE*, vol. 94, no. 6, pp. 1089-1120, 2006.

5. S. Kundu and A. Sreedhar, *Nanoscale CMOS VLSI Circuits: Design for Manufacturability*, McGraw-Hill, New York, USA, 2010.

6. G. Bouesse, G. Sicard, A. Baixas and M. Renaudin, "Quasi delay insensitive asynchronous circuits for low EMI," *Proceedings of 4th International Workshop on Electromagnetic Compatibility of Integrated Circuits*, pp. 27-31, 2004.

7. L.A. Plana, P.A. Riocreux, W.J. Bainbridge, A. Bardsley, S. Temple, J.D. Garside and Z.C. Yu, "SPA – a secure amulet core for smartcard applications," *Microprocessors and Microsystems*, vol. 27, no. 9, pp. 431-446, 2003.

8. M. Renaudin and Y. Monnet, "Asynchronous design: fault robustness and security characteristics," *Proceedings of 12th IEEE International On-Line Testing Symposium on*, pp. 1-4, 2006.

9. J. Sparsø and S. Furber (Eds.), *Principles of Asynchronous Circuit Design: A Systems Perspective*, Kluwer Academic Publishers, 2001.

10. A.J. Martin, "Can asynchronous techniques help the SoC designer?," *Proceedings of IFIP International Conference on VLSI-SoC*, pp. 7-11, 2006.

11. K.-L. Chang, J.S. Chang, B.-H. Gwee and K.-S. Chong, "Synchronous-logic and asynchronous-logic 8051 microcontroller cores for realizing the internet of things: a comparative study on dynamic voltage scaling and variation effects," *IEEE Journal on Emerging and Selected Topics in Circuits and Systems*, vol. 3, no. 1, pp. 23-34, 2013.

12. A.J. Martin, "The limitation to delay-insensitivity in asynchronous circuits," *Proceedings of 6th MIT Conference on Advanced Research in VLSI*, pp. 263-278, 1990.

13. A.J. Martin and P. Prakash, "Asynchronous nano-electronics: preliminary investigation," *Proceedings of 14th IEEE International Symposium on Asynchronous Circuits and Systems*, pp. 58-68, 2008.

14. D.E. Muller and W.S. Bartky, "A theory of asynchronous circuits," *Proceedings of International Symposium on the Theory of Switching*, Part I, pp. 204-243, Harvard University Press, 1959.

15. B. Bose, "On unordered codes," *IEEE Transactions on Computers*, vol. 40, no. 2, pp. 125-131, 1991.

16. S.J. Piestrak and T. Nanya, "Towards totally self-checking delay-insensitive systems," *Proceedings of 25th International Symposium on Fault-Tolerant Computing*, pp. 228-237, 1995.

17. M.T. Moreira, R.A. Guazzelli and N.L.V. Calazans, "Return-to-one protocol for reducing static power in C-elements of QDI circuits employing m-of-n codes," *Proceedings of 25th Symposium on Integrated Circuits and Systems Design*, pp. 1-6, 2012.

18. C.L. Seitz, "System Timing," in *Introduction to VLSI Systems*, C. Mead and L. Conway (Eds.), pp. 218-262, Addison-Wesley, Reading, MA, USA, 1980.

19. P. Balasubramanian and D.A. Edwards, "Efficient realization of strongly indicating function blocks," *Proceedings of IEEE Computer Society Annual Symposium on VLSI*, pp. 429-432, 2008.

20. P. Balasubramanian and D.A. Edwards, "A new design technique for weakly indicating function blocks," *Proceedings of 11th IEEE Workshop on Design and Diagnostics of Electronic Circuits and Systems*, pp. 116-121, 2008.

21. P. Balasubramanian, "A robust asynchronous early output full adder," *WSEAS Transactions on Circuits and Systems*, vol. 10, no. 7, pp. 221-230, 2011.

22. P. Balasubramanian and N.E. Mastorakis, "Global versus local weak-indication self-timed function blocks – a comparative analysis," *Proceedings of 10$^{th}$ International Conference on Circuits, Systems, Signal and Telecommunications*, pp. 86-97, 2016.

23. P. Balasubramanian and N.E. Mastorakis, "Analyzing the impact of local and global indication on a self-timed system," *Proceedings of 5$^{th}$ European Computing Conference*, pp. 85-91, 2011.

24. P. Balasubramanian and N.E. Mastorakis, "Timing analysis of quasi-delay-insensitive ripple carry adders – a mathematical study," *Proceedings of 3$^{rd}$ European Conference of Circuits Technology and Devices*, pp. 233-240, 2012.

25. P. Balasubramanian, C. Jacob Prathap Raj, S. Anandhi, U. Bhavanidevi and N.E. Mastorakis, "Mathematical modeling of timing attributes of self-timed carry select adders," *Proceedings of 4$^{th}$ European Conference of Circuits Technology and Devices*, pp. 228-243, 2013.

26. W.B. Toms, "Synthesis of quasi-delay-insensitive datapath circuits," Ph.D. thesis, School of Computer Science, The University of Manchester, UK, 2006.

27. P. Balasubramanian and N.E. Mastorakis, "QDI decomposed DIMS method featuring homogeneous/heterogeneous data encoding," *Proceedings of International Conference on Computers, Digital Communications and Computing*, pp. 93-101, 2011.

28. P. Balasubramanian and R. Arisaka, "A set theory based factoring technique and its use for low power logic design," *International Journal of Electrical, Computer and Systems Engineering*, vol. 1, no. 3, pp. 188-198, 2007.

29. P. Balasubramanian, K. Prasad and N.E. Mastorakis, "Robust asynchronous implementation of Boolean functions on the basis of duality," *Proceedings of 14$^{th}$ WSEAS International Conference on Circuits*, pp. 37-43, 2010.

30. P. Balasubramanian, Comments on "Dual-rail asynchronous logic multi-level implementation", *Integration, the VLSI Journal*, vol. 52, no. 1, pp. 34-40, 2016.

31. P. Balasubramanian, Critique of "Asynchronous logic implementation based on factorized DIMS", *arXiv preprint* arXiv:1711.02333, 2017.

32. V.I. Varshavsky (Ed.), *Self-Timed Control of Concurrent Processes: The Design of Aperiodic Logical Circuits in Computers and Discrete Systems*, Chapter 4: Aperiodic Circuits, pp. 77-85, (Translated from the Russian by Alexandre V. Yakovlev), Kluwer Academic Publishers, 1990.

33. P. Balasubramanian, R. Arisaka and H.R. Arabnia, "RB_DSOP: a rule based disjoint sum of products synthesis method," *Proceedings of 12th International Conference on Computer Design*, pp. 39-43, 2012.

34. P. Balasubramanian and D.A. Edwards, "Self-timed realization of combinational logic," *Proceedings of 19th International Workshop on Logic and Synthesis*, pp. 55-62, 2010.

35. P. Balasubramanian and N.E. Mastorakis, "A set theory based method to derive network reliability expressions of complex system topologies," *Proceedings of Applied Computing Conference*, pp. 108-114, 2010.

36. P. Balasubramanian, D. Maskell and N. Mastorakis, "Low power robust early output asynchronous block carry lookahead adder with redundant carry logic," *Electronics*, vol. 7, no. 10, Article #243, pp. 1-21, 2018.

37. C. Brej, "Early output logic and anti-tokens," Ph.D. thesis, School of Computer Science, The University of Manchester, UK, 2006.

38. Synopsys SAED_EDK32/28_CORE Databook, Revision 1.0.0, 2012.

39. P. Balasubramanian, D.L. Maskell and N.E. Mastorakis, "Speed and energy optimized quasi-delay-insensitive block carry lookahead adder," *PLOS ONE*, vol. 14, no. 6, Article ID e0218347, pages 27, 2019.

40. P. Balasubramanian and S. Yamashita, "Area/latency optimized early output asynchronous full adders and relative-timed ripple carry adders," *SpringerPlus*, vol. 5:440, no. 1, pp. 1-26, 2016.

41. P. Balasubramanian and K. Prasad, "Early output hybrid input encoded asynchronous full adder and relative-time ripple carry adder," *Proceedings of 14th International Conference on Embedded Systems, Cyber-physical Systems, and Applications*, pp. 62-65, 2016.

42. P. Balasubramanian, "An asynchronous early output full adder and a relative-timed ripple carry adder," *WSEAS Transactions on Circuits and Systems*, vol. 15, Article #12, pp. 91-101, 2016.

43. K.S. Stevens, R. Ginosar and S. Rotem, "Relative timing," *IEEE Transactions on VLSI Systems*, vol. 11, no. 1, pp. 129-140, 2003.

44. N.P. Singh, "A design methodology for self-timed systems," *M.Sc. dissertation*, Massachusetts Institute of Technology, USA, 1981.

45. J. Sparsø and J. Staunstrup, "Delay-insensitive multi-ring structures," *Integration, the VLSI Journal*, vol. 15, no. 3, pp. 313-340, 1993.

46. B. Folco, V. Bregier, L. Fesquet and M. Renaudin, "Technology mapping for area optimized quasi delay insensitive circuits," *Proceedings of IFIP 13th International Conference on VLSI-SoC*, pp. 146-151, 2005.

47. P. Balasubramanian and D.A. Edwards, "A delay efficient robust self-timed full adder," *Proceedings of IEEE 3rd International Design and Test Workshop*, pp. 129-134, 2008.

48. P. Balasubramanian, "A latency optimized biased implementation style weak-indication self-timed full adder," *Facta Universitatis, Series: Electronics and Energetics*, vol. 28, no. 4, pp. 657-671, 2015.

49. P. Balasubramanian and K. Prasad, "Asynchronous early output dual-bit full adders based on homogeneous and heterogeneous delay-insensitive data encoding," *WSEAS Transactions on Circuits and Systems*, vol. 16, Article #8, pp. 64-73, 2017.

50. P. Balasubramanian and K. Prasad, "Latency optimized asynchronous early output ripple carry adder based on delay-insensitive dual-rail data encoding," *International Journal of Circuits, Systems and Signal Processing*, vol. 11, pp. 65-74, 2017.

51. P. Balasubramanian, "Asynchronous ripple carry adder based on area optimized early output dual-bit full adder," *arXiv preprint* arXiv:1807.09762, 2018.

52. P. Balasubramanian, "Asynchronous carry select adders," *Engineering Science and Technology, an International Journal*, vol. 20, no. 3, pp. 1066-1074, 2017.

53. P. Balasubramanian, D.A. Edwards and W.B. Toms, "Self-timed section-carry based carry lookahead adders and the concept of alias logic," *Journal of Circuits, Systems, and Computers*, vol. 22, no. 4, pp. 1350028-1 – 1350028-24, 2013.

54. P. Balasubramanian, D. Dhivyaa, J.P. Jayakirthika, P. Kaviyarasi and K. Prasad, "Low power self-timed carry lookahead adders," *Proceedings of 56th IEEE International Midwest Symposium on Circuits and Systems*, pp. 457-460, 2013.

55. P. Balasubramanian, C. Dang, D.L. Maskell and K. Prasad, "Asynchronous early output section-carry based carry lookahead adder with alias carry logic," *Proceedings of IEEE 30th International Conference on Microelectronics*, pp. 293-298, 2017.

56. P. Balasubramanian and D.A. Edwards, "Dual-sum single-carry self-timed adder designs," *Proceedings of IEEE Computer Society Annual Symposium on VLSI*, pp. 121-126, 2009.

57. P. Balasubramanian and D.A. Edwards, "Heterogeneously encoded dual-bit self-timed adder," *Proceedings of IEEE Ph.D. Research in Microelectronics and Electronics Conference*, pp. 120-123, 2009.

58. P. Balasubramanian and C. Dang, "A comparison of quasi-delay-insensitive asynchronous adder designs corresponding to return-to-zero and return-to-one handshaking," *Proceedings of 60th IEEE International Midwest Symposium on Circuits and Systems*, pp. 1192-1195, 2017.

59. P. Balasubramanian, "Comparative evaluation of quasi-delay-insensitive asynchronous adders corresponding to return-to-zero and return-to-one handshaking," *Facta Universitatis, Series: Electronics and Energetics, Invited Paper*, vol. 31, no. 1, pp. 25-39, 2018.

60. P. Balasubramanian and D.A. Edwards, "Power, delay and area efficient self-timed multiplexer and demultiplexer designs," *Proceedings of IEEE 4th International Conference on Design and Technology of Integrated Systems in Nanoscale Era*, pp. 173-178, 2009.

61. P. Balasubramanian, D.A. Edwards and W.B. Toms, "Redundant logic insertion and latency reduction in self-timed adders," *VLSI Design*, vol. 2012, Article ID 575389, pp. 1-13, 2012.

62. P. Balasubramanian and N. Mastorakis, "Performance comparison of carry-lookahead and carry-select adders based on accurate and approximate additions," *Electronics*, vol. 7, no. 12, Article #369, pages 12, 2018.

63. P. Balasubramanian, "Performance comparison of some synchronous adders," *arXiv preprint* arXiv:1810.01115, 2018.